\documentclass[rnote]{aa}

\usepackage{graphicx}
\usepackage{txfonts}
\usepackage{hyperref}
\usepackage{natbib}
\bibpunct{(}{)}{;}{a}{}{,} 

\usepackage{epsfig}
\usepackage{epstopdf}                   
\usepackage{multirow}                   
\usepackage{threeparttable}             
\usepackage{tabularx}                   
\usepackage{booktabs}                   

\begin{document}

\title{SiO: Not the perfect outflow tracer
  \thanks{Based on observations carried out with the PdBI and the IRAM 30\,m telescope}\fnmsep
  \thanks{Appendices are available in electronic form at \url{http://www.aanda.org}}
  }
  
   \subtitle{Outflow studies of the massive star formation region IRAS 19410+2336}

   \author{F.~Widmann
          \inst{1} \thanks{E-mail: f.widmann@stud.uni-heidelberg.de}
          \and
          H.~Beuther
          \inst{1}
          \and
          P.~Schilke
          \inst{2}
          \and
          T.~Stanke
          \inst{3}
          }
   \institute{Max Planck Institute for Astronomy, K\"onigstuhl 17,
              69117 Heidelberg, Germany
              \and
              I. Physikalisches Institut, University of Cologne, Z\"ulpicher Strasse 77,                               50937 K\"oln, Germany
              \and
              European Southern Observatory, Karl-Schwarzschild-Stra\ss e 2, 85748 Garching bei M\"unchen, Germany}
             
   \date{Received 15 May 2015 / Accepted 24 February 2016}

\abstract
{}
{Previous observations of the young massive star formation region IRAS 19410+2336 have revealed strong outflow activity with several interacting outflows. We aim to get a better understanding of the outflow activity in this region by observing the SiO and H$^{13}$CO$^+$ emission with high angular resolution. SiO is known to trace shocked gas, which is often associated with young energetic outflows. With the H$^{13}$CO$^+$ data, we intend to better understand the distribution of the quiescent gaseous component of the region.}
{The SiO observations in the J=2-1 v=0 transition and H$^{13}$CO$^+$ J=1-0 observations were performed by the Plateau de Bure Interferometer, combined with IRAM 30 m single-dish observations, in order to get the missing short-spacing information. We complement this new high-resolution observation with earlier CO and H$_2$ data.}
{The SiO observations do not trace the previously in CO and H$_2$ identified outflows well. Although we identify regions of highly increased SiO abundance indicative of shock interaction, there are hardly any bipolar structures in the data. The southern part of the region, which exhibits strong H$_2$ emission, shows almost no SiO. The CO and SiO data show only weak similarities, and the main SiO emission lies between the two dominating dust clumps of the region.}
{Most SiO emission is likely to be a result of high-velocity shocks due to protostellar jets. However, this does not explain all the emission features and additional effects; for example, colliding gas flows at the interface of the two main regions may play an important role in the origin of the emission. The present SiO data show that several different effects can influence SiO emission, which makes the interpretation of SiO data more difficult than often assumed.}
\keywords{Stars: formation -- Stars: massive -- Stars: individual: IRAS 19410+2336 -- Molecular data -- ISM: jets and outflows}


\maketitle
\section{Introduction}\label{intro}
The formation of massive stars is the subject of strong debate, and there are several theoretical approaches that are able to explain it. One of the main theories is the \textit{Core Accretion Model}. It assumes that the processes in high-mass star formation are similar to the processes in low-mass star formation. The main difference for high-mass stars is given by massive turbulent cores, which lead to higher accretion rates on the protostars and massive disks around them \citep[see][]{Krumholz2006,Mckee2003}. Another approach is the so-called \textit{Competitive Accretion Model}, which is based on the observation that massive stars form in dense clusters. In this scenario, the cloud fragments into many low-mass cores, which form protostars. Since these cores are close together in the dense center of the cloud, they act together and produce a large scale gravitational potential. This leads to the accretion of material from a larger area than would be possible for a single protostar. In the following, the cores in the dense center compete to accrete mass from this area \citep[see][]{Bonnell2007,Bonnell2008,Zinnecker2007}. Both models expect disks and jets around the protostars, but in different forms, because the clustered formation of protostars in the competitive accretion would most likely lead to less collimated outflows. The core accretion model, on the other hand, is a scaled-up version of low-mass star formation and therefore expects massive collimated outflows, similar to the ones found in low-mass star formation \citep{Tan2014}. 

Since direct observations of the innermost regions around the central disks and cores are very difficult, a study of the jets and outflows is helpful for understanding what is happening during the formation of the stars. A tracer that is often used to detect the jets and outflows is SiO, because the depletion of Si in quiescent gas is high and most of the Si can be found in silicates, bound to dust grains, so the Si abundance in quiet regions is low. However, dynamical processes, such as shocks, are able to release the silicon from the dust grains. The silicon can be directly present as SiO on the dust grains or is transformed by successive gas-phase chemistry from Si into SiO \citep[see Section \ref{dis} for further discussion]{Schilke1997,Gusdorf2008a,Gusdorf2008b}.
One therefore assumes that SiO shows the shocks due to jets or outflows directly and is less influenced by environmental gas, such as more common molecules like CO \citep[see, e.g.,][]{Codella2007,Lee2007}. This was proven by several SiO observations toward massive star formation regions, where the protostellar jets were revealed by the SiO emission \citep[e.g.,][]{Cesaroni1999,Beuther2002,Leurini2013,Codella2013}.

However, there are also studies in which the origin of the SiO emission is more ambiguous, since it does not show a clear jet or outflow structure. An example is the observation of the massive star formation region W43 by \citet{NguyenLuong2013}. The observation revealed large scale SiO emission, which is probably not a result of jets due to star formation but of low-velocity shocks due to colliding flows. Another example is the young outflow region L1448-mm, which has revealed narrow ($\Delta$v $\approx$ 0.5 km s$^{-1}$) SiO lines at ambient velocities, which could in turn be emission from decelerated postshock material \citep{Gusdorf2008a} or produced by magnetic shock precursors \citep{Jimenez2004}. In addition, there are still discussions of the time evolution of the SiO emission. In this context, \citet{Jimenez2009,Jimenez2011} investigated the time evolution of the shock propagation and \citet{LopezSepulcre2011} and \citet{Leurini2014} the time dependence in jet activity.  They came to the conclusion that the SiO emission due to massive jet-driven outflows can strongly change on comparably short timescales.

\begin{figure}
\centering\includegraphics[width=0.45\textwidth]{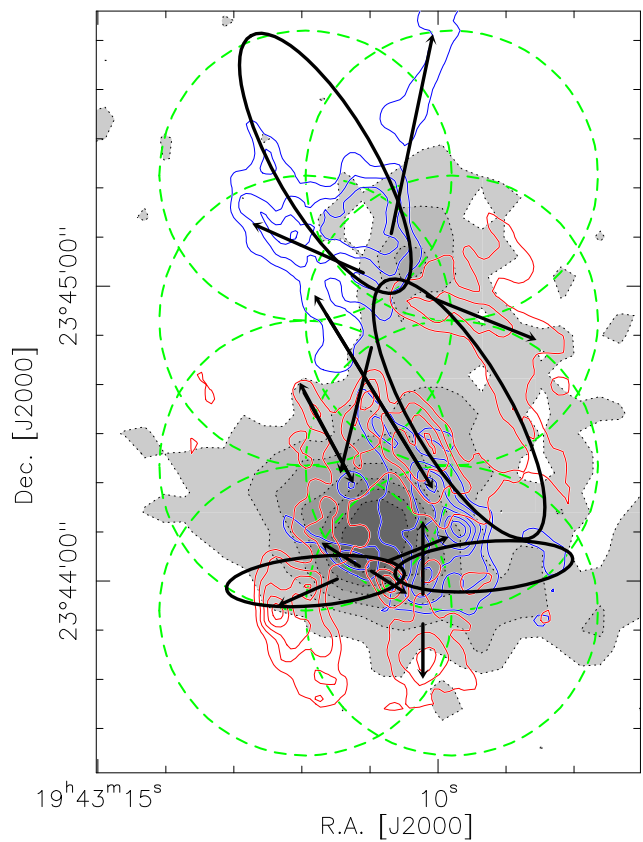}
\caption{Proposed outflows by \citet{Beuther2003} shown by ellipses and arrows. The proposal of the outflows is based on the line wing data of CO (1-0), as shown in the figure in blue and red contours. The blue contours show the line emission from CO, integrated from 2 to 18 km s$^{-1}$, and the red ones show the emission integrated from 26 to 47 km s$^{-1}$.
The contour levels are in both cases from 2 to 20 Jy beam$^{-1}\cdot$ km s$^{-1}$ in steps of 3 Jy beam$^{-1}\cdot$ km s$^{-1}$, in short 2(3)20 Jy beam$^{-1}\cdot$ km s$^{-1}$. The background shows the 1.2 mm continuum data from \citet{Beuther2002} in gray scale with dotted contours (the contour levels are 100(100)700 mJy beam$^{-1}$).
The green dotted circles show the mosaic coverage of our new interferometric observations.}
\label{outflows}
\label{coverage}
\end{figure}

The present work is based on observations of the region IRAS 19410+2336, which is a young massive star formation region in an early evolutionary state. It has been studied a lot, and the main characteristics of the region are well known. The distance to the region is 2.16 kpc \citep{Xu2009}, and it has a rest velocity of 22.4 km s$^{-1}$ \citep{Sridharan2002} and an integrated bolometric luminosity of around $10^4$ L$_\odot$ \citep {Martin2008}. The total gas mass of this region is calculated by \citet{Beuther2002} and \citet{Beuther2005} to be  510 M$_\odot$. The mean temperature was investigated by \citet{Rodon2012} with several known gas temperature tracers, such as H$_2$CO and CH$_3$CN, with the result of an average temperature of 40 $\pm$ 15 K and a temperature of 80 $\pm$ 40 K for the dense cores. Early observations of the continuum of IRAS 19410+2336 were conducted by \citet{Beuther2002} and \citet{Beuther2004}. Of special interest concerning the formation of massive stars is the outflow study from \citet{Beuther2003}. They observed the 2.6 mm continuum and the CO (1-0) line, using the IRAM Plateau de Bure Interferometer and the IRAM 30 m telescope. Their study also included observations of the H$_2$ line at 2.12 $\mu$m, using the Calar Alto 3.5 m telescope. They found an intense outflow activity with strong evidence of possibly nine, but at least seven bipolar outflows. The proposed outflows are shown in Figure \ref{outflows}.

However, their study was based on observations of the CO (1-0) line. CO emission with a high velocity shift can be used to trace outflows and shocks. But CO is very common in the interstellar medium, and the CO (1-0) transition is already excited at low temperature. Therefore the emission is strongly influenced by environmental gas and this line is not considered to be an ideal outflow tracer. The aim of the new observation was to get a better understanding of the outflow activity, using SiO as a well-known shock tracer, which should be less influenced by the environmental gas and therefore show the jets and outflows better.

\section{Observations}\label{observations} 
The observations were done in several steps. The interferometric data were taken with the Plateau de Bure Interferometer in France \citep[PdBI, see][]{Guilloteau1992} and were merged with the single-dish data from the IRAM 30 m telescope at Pico Veleta, Spain.  The CO and H$_2$ data were already available as  calibrated datasets by \citet{Beuther2003}, as well as a 1.2 mm continuum map by \citet{Beuther2002}. 

The region was observed with the PdBI in May 2003 in the D Array with five antennas, and in December 2003 in the C Array with six antennas. The baselines ranged from 24 m to 82 m in the D-Array, and from 48 m to 229 m in the C Array. It was observed simultaneously at 3 mm with receiver 1 and at 1 mm with receiver 2.  Bad weather conditions meant that the quality of the 1 mm observation is poor so the data were not used in the following analysis. Receiver 1 was tuned to 86.847 GHz toward the SiO (2-1) line. Simultaneously, the H$^{13}$CO$^+$(1-0) line at 86.754 GHz was observed with another correlator unit. In addition, two 320 MHz correlators were used to get a continuum measurement with a large bandwidth. The primary beam of the array at 3 mm is 59'' making it impossible to observe the whole region at once. As shown in Fig. \ref{coverage}, a mosaic of eight fields was observed to cover the whole area.

To remove temporal fluctuations of the amplitude and the phase, the quasars 1923+210 and 2023+336 were used. To get the total value of the amplitude, the two quasars MWC349 and 3C454 were observed and another quasar, 0420-014, was observed for the bandpass calibration. The frequency resolution of the used correlators is 40 kHz, which means a velocity resolution of 0.14 km s$^{-1}$. For further analysis, the data were smoothed into velocity channels of 1 km s$^{-1}$.

The observations with the IRAM 30m telescope were carried out in August 2003 in the on-the-fly mode. The SiO (2-1) and the H$^{13}$CO$^+$ (1-0) lines were observed simultaneously with a frequency resolution of 40 kHz. This is equivalent to a velocity resolution of 0.28 km s$^{-1}$. Also, the single-dish data are smoothed to 1 km s$^{-1}$. 

The two data sets were merged, using the GILDAS\footnote{\url{http://www.iram.fr/IRAMFR/GILDAS}} software. The final resolution of the 3 mm continuum is 4.28 x 2.68 arcseconds at a position angle of 32 degree and for the lines 4.80 x 3.06 arcseconds at a position angle of 31 degrees. The rms of the final data is 4.8 x 10$^{-4}$ Jy beam$^{-1}$ for the continuum and 1.7 x 10$^{-2}$ Jy beam$^{-1}$ for the line data.

\section{Observational results}
\begin{figure*}
\centering
\includegraphics[width=0.9\textwidth]{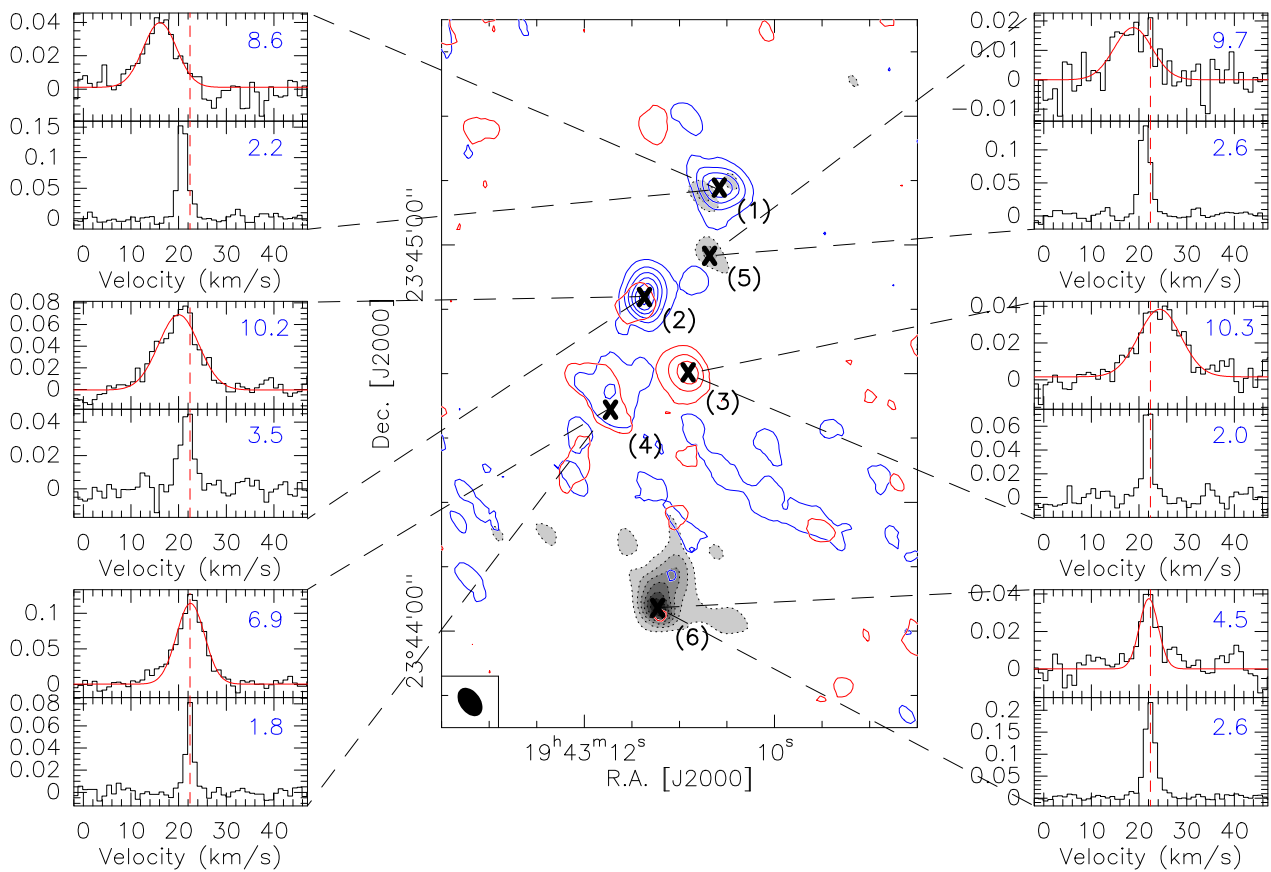}
\caption{Central panel: Integrated intensity of the blue and redshifted wing of the SiO (2-1) emission. In blue contours the blueshifted wing is shown, integrated from 6 km s$^{-1}$ to 19 km s$^{-1}$ (levels 0.2(0.2)2 Jy beam$^{-1}\cdot$ km s$^{-1}$). The red contours show the redshifted wing, integrated from 25 km s$^{-1}$ to 38 km s$^{-1}$ (levels 0.2(0.2)2 Jy beam$^{-1}\cdot$ km s$^{-1}$). In the background, the 3 mm continuum data in gray and dotted contours (levels 1.5(1.5)10 Jy beam$^{-1}$) are shown. Left and right sides: Spectra at different points.  For each point, the upper spectrum is from SiO and the lower one from H$^{13}$CO$^+$, each in the units of Jy beam$^{-1}$. The red dotted line shows the rest velocity of 22.4 km s$^{-1}$ and the blue number at the upper right of each spectrum the FWHM of the fitted Gaussian in km s$^{-1}$. In the SiO spectra, the fitted Gaussians are shown with red lines.}
\label{spektren1}
\end{figure*}

The main goal of the observation was to gain information about the outflow activity in the region. Therefore, the line wings of the SiO emission are important to study. The line wings are the part of the emission which is a result of moving material along the line of sight, for example due to outflows. As indication for the line width of the ambient gaseous component, we used the width of H$^{13}$CO$^+$ line, averaged over the whole area, which is on average 6 km s$^{-1}$ for the full width at zero power (FWZP) and 2.7 km s$^{-1}$ for the full width half maximum (FWHM). This means that the emission of the line wings starts approximately at the rest velocity of 22.4 km s$^{-1}$ $\pm$ 3 km s$^{-1}$. With this information and the visual inspection of the SiO emission, we came to the conclusion that the emission between 6 and 19 km s$^{-1}$ best shows the blueshifted wing and the emission between 25 and 38 km s$^{-1}$ the redshifted one.

Figure \ref{spektren1} (central panel) presents the 3 mm continuum, as well as the line-wing emission of the SiO emission. The continuum structure fits older observations shown in Fig. \ref{coverage}
well \citep[see also][]{Beuther2002,Beuther2004,Rodon2012}. The interferometric observation filters out large scale structure, but the comparison to older interferometric observation at 2.6mm by \citet{Beuther2003} shows very good agreement. Also, the existence of the strong SiO emission is a clear hint of ongoing shock activity. But there are unexpected results. The majority of the protostars are expected to form in the densest areas of the region, which are shown by the strongest continuum emission. These are also the areas where \citet{Beuther2003} found most of the high-velocity CO emission. Therefore, one would expect most shock and jet activity, and consequently most SiO emission in the same areas or the close environment of the main continuum emission and high velocity CO emission. However, most of the SiO emission occurs between the two large scale continuum sources and in the northern part of the observed region, whereas only weak emission appears around the southern continuum sources. Another observation from Fig. \ref{spektren1} is that there are no clear bipolar structures. 

\subsection{Emission lines}\label{linewidth}
To get a better understanding of the data, it is helpful to look at the spectra at different points, which are shown in Fig. \ref{spektren1}. These are the points of the strongest blue- (Points 1 and 2) and redshifted (Points 3 and 4) SiO emission and also the centers of the two massive gas clumps (Point 5 and 6). In each spectrum, a fitted  Gaussian shows the key characteristics of the emission line. As we expect a lot of emission due to shocks, it is unlikely that a Gaussian fits\ the emission very well. However, we can use it to derive some characteristics of the emission line, such as the peak velocity and an approximation of the FWHM. For more information on the Gaussians, see Table \ref{table_spek} in the appendix. 
The FWHM of the H$^{13}$CO$^+$ lines are between 1.8 km s$^{-1}$ and 3.5 km s$^{-1}$. This confirms that H$^{13}$CO$^+$ only traces the dense gas without any significant linewings and does not show shock activity \citep[e.g.,][]{Vasyunina2011}. The thermal line width at 40 K of the H$^{13}$CO$^+$ line is smaller than the observed one ($\Delta$v $\approx$ 0.22 km s$^{-1}$) because the line is broadened due to turbulent motion in the cloud. \citet{Beuther2007a} measured line widths between 1.8 and 4.4 km s$^{-1}$ for the H$^{13}$CO$^+$ line for massive infrared dark clouds, which fits our observations.

In comparison to the H$^{13}$CO$^+$ lines, the lines from SiO are much broader. Interestingly, this broader emission is often fit relatively well by Gaussian profiles, but we also identify clear line wing emission in several positions (especially Points 2 \& 3). We derived Gaussian FWHM values between 4.5 km s$^{-1}$ and 11.3 km s$^{-1}$, as shown in Table \ref{table_spek}. The line widths therefore lie in a range that is also observed by other studies, since the SiO lines often have widths of 5 km s$^{-1}$ to 20 km s$^{-1}$ \citep[e.g.,][]{Gusdorf2008a,Motte2007}. In the spectra averaged over the whole area, we derived a FWZP of approximately 18 km $^{-1}$. This also lies in the expected range, but is a rather low value for a massive star formation region. The typical value for the FWZP lies between 10 km up to 90 km \citep[e.g.,][]{NguyenLuong2013}.

The line widths therefore support the statement that SiO traces  dynamical gas in jets and outflows. Additionally, most of the lines in the  spectra peak below or above the rest velocity and are represented by a Gaussian, but with an additional signal in the line wings, which again indicates there are jets or outflows. \\
The spectra at Points 4 and 6 are remarkable because their line peaks are approximately at rest velocity. The spectrum at Point 6 is an exception in a way, since it is extracted at the densest part of the cloud, and the signal is within the six sigma range, which is relatively weak, while the emission at Point 4 is over ten sigma strong. This SiO peak clearly lies outside of the dust emission and is the only one of the strong emission peaks that is not significantly red- or blueshifted, but it lies exactly at rest velocity. Furthermore, this emission line has a width of 6.9 km s$^{-1}$, one of the narrowest of the extracted spectra.

\subsection{Comparison with older data}
\begin{figure*}
\centering
\includegraphics[width=0.95\textwidth]{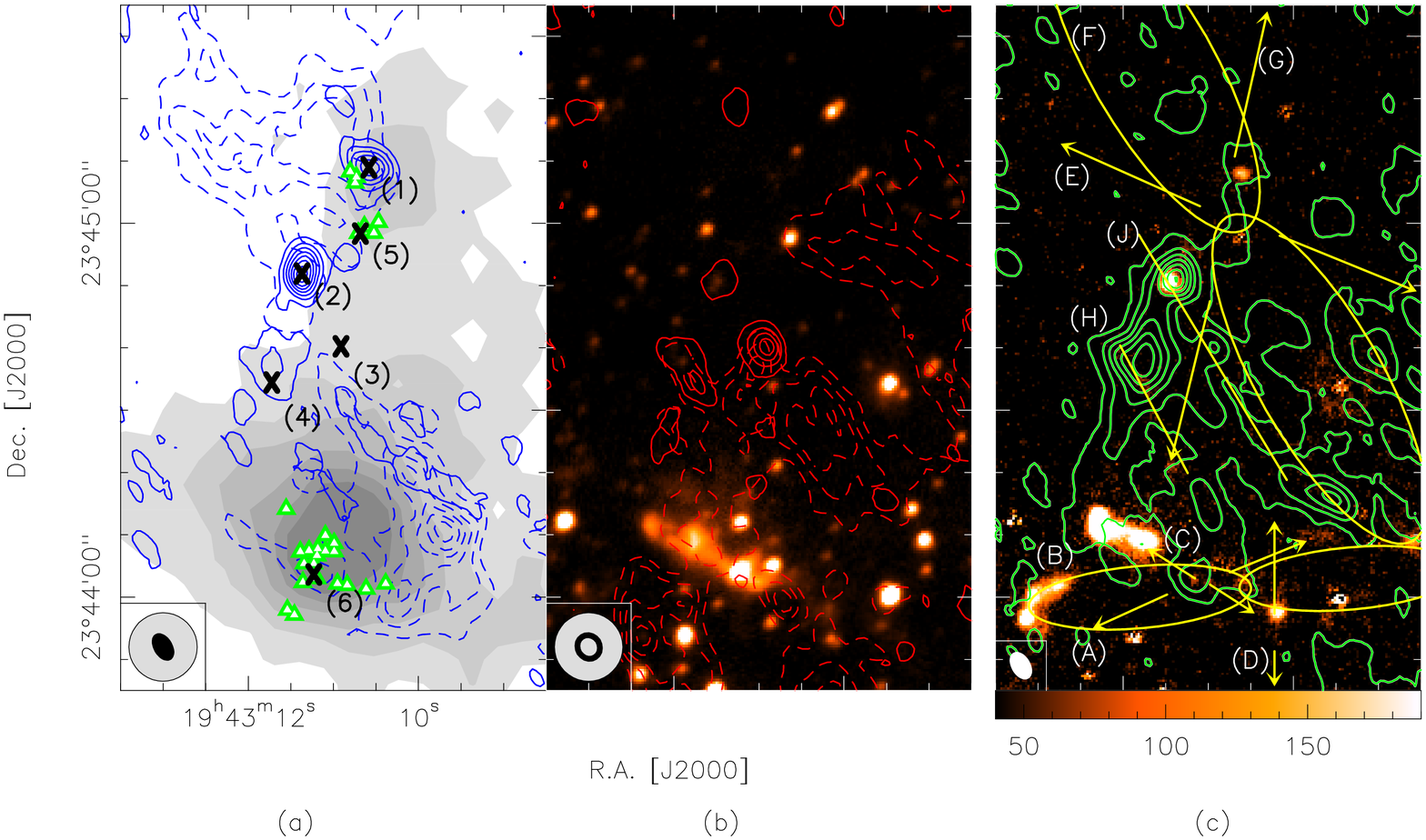}
\caption{(a) \& (b): Comparison of the SiO (2-1) and CO (1-0) data. The left image shows the blueshifted wings of both datasets, with CO in the range 2-18 km s$^{-1}$ with dotted contours and the SiO 6-19 km s$^{-1}$ in solid contours. The middle image shows the redshifted counterparts, SiO in solid contours in the area 25-38 km s$^{-1}$, and CO in dotted contours from 26 to 47 km s$^{-1}$. The contour levels of the SiO emission are 0.2(0.2)2 Jy beam$^{-1}\cdot$ km s$^{-1}$ and 2(3)20 Jy beam$^{-1}\cdot$ km s$^{-1}$ for the CO emission. The background of the left image shows the 1.2 mm continuum data from \citet{Beuther2002} and the background of the middle image the K-Band emission observed by \citet{Beuther2002c}. The green triangles in the left image indicate mm-sources from \citet{Rodon2012} and the crosses and numbers label the same points as in Fig. \ref{spektren1}. The resolution of the continuum data is shown by the gray circle in the lower left corner, and the black ellipses show the resolution of the SiO (filled) and the CO data (open). (c): SiO emission at ambient velocity in green contours (integrated from 19 to 25 km s$^{-1}$, contour levels 0.2(0.4)2 Jy beam$^{-1}\cdot$ km s$^{-1}$) with the H$_2$ emission in the background and the proposed outflows in yellow lines, arrows, and ellipses. CO and H$_2$ data, as well as the proposed outflows, are from \citet{Beuther2003}.}
\label{sio_co}
\end{figure*}

Now we analyze the new data with respect to already existing observations from other outflow and shock tracers such as CO and H$_2$ \citep{Beuther2003}. CO traces the environmental gas, and therefore the high-velocity emission shows the entrained outflow gas \citep{Arce2007}. Both SiO and the high velocity component of CO are well known outflow tracers; however, the SiO emission should show the jets and outflows clearer than CO. In Fig. \ref{sio_co}, the lefthand image shows the blueshifted wing of the SiO and CO emission line, whereas the middle image shows the redshifted wing. 

At first sight, the two molecules seem to have a very different distribution. Actually, there are some SiO structures that correlate with the CO (e.g., around Point 4 and the long structure under Point 3, which is visible in the blueshifted wing), but they are quite rare. Since the CO traces not only the gas from the outflows but also the environmental gas, the structures in the SiO emission are expected to be smaller and less wide. This is clearly observable in our data. The large bipolar structures that are clearly present in the CO data, in particular, are largely missing in the SiO emission. Again, it is remarkable that the southern part lacks SiO emission, although there is plenty of high-velocity CO emission. One can also see that there are some structures that are clearly traced in SiO, but have no counterpart in CO. \\
Because we found a lot of SiO signal at ambient velocity, the righthand image of Fig. 3 shows a map of the SiO emission between 19 and 25 km s$^{-1}$. This map is combined with the H$_2$ data at 2.12 $\mu$m and the proposed outflows by \citet{Beuther2003}. The H$_2$ emission traces shock activity and can deliver some information about the different kinds of shocks, which is discussed below. One can see that there is plenty of H$_2$ activity in the region. The two strongest H$_2$ peaks in the northern part seem to be directly correlated with SiO emission.
Also, the SiO emission at ambient velocity shows some agreement with the proposed outflows in the northern part; for example, the outflows J and H can be identified with SiO emission, as well as the southern part of outflow F.
However, since the SiO emission in the south is very weak, there is no accordance with the four proposed outflows in this area. Furthermore, most of the H$_2$ emission is located in the southern part, where several strong H$_2$ features with jet-like morphological structures are visible, which are also associated with the proposed outflows. It is surprising that especially these H$_2$ structures, which are indicative of strong shocks, have no SiO counterparts.

\subsection{SiO abundance}
For a better understanding of the SiO distribution, Fig. \ref{abundance} shows the N(SiO)/N(H$_2$) abundance of the region. For the calculations, see Appendix \ref{A.abundance}.
\begin{figure}
         \centering
        \includegraphics[width=0.4\textwidth]{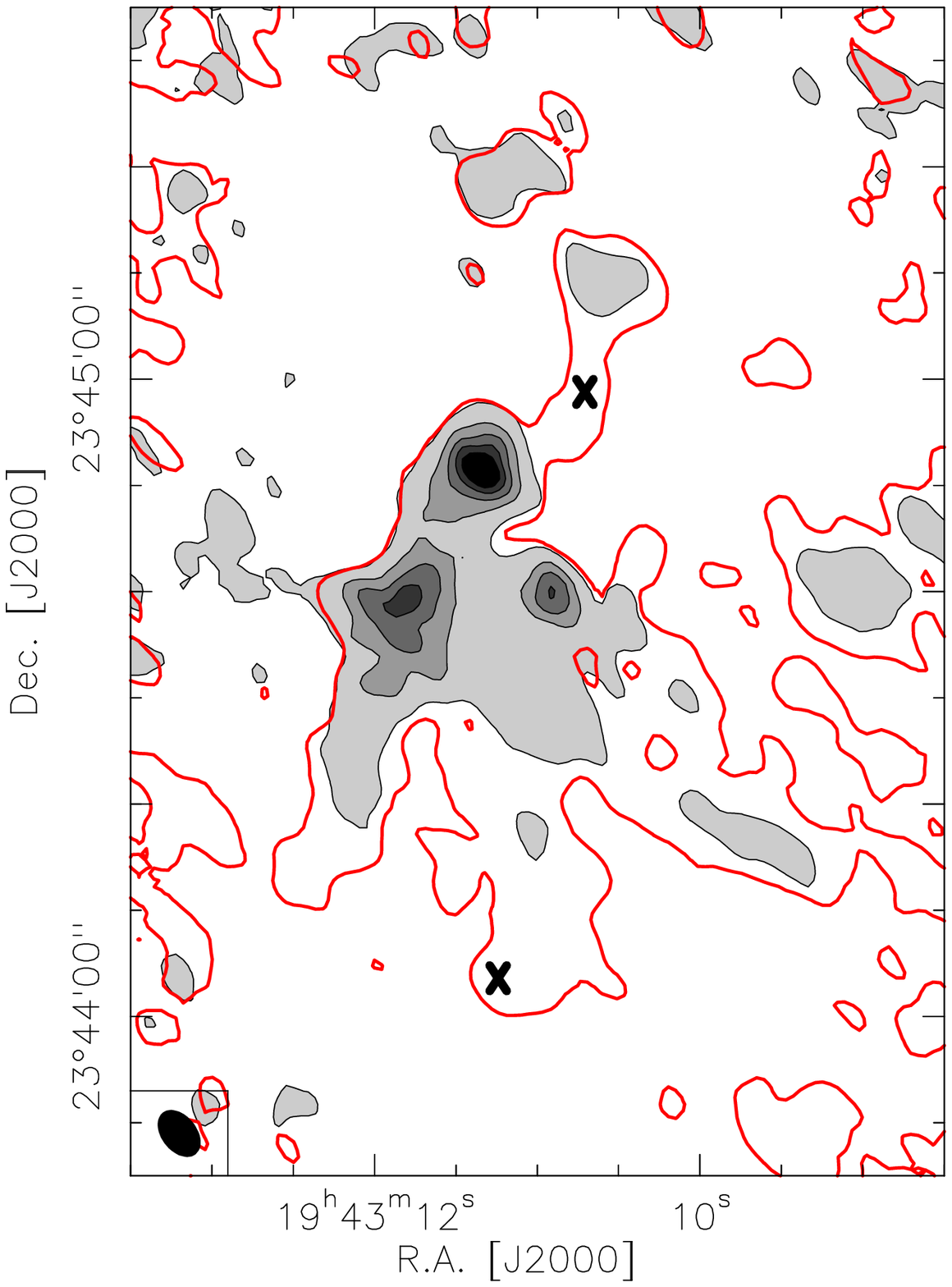}
        \caption{Abundance map of SiO, computed by comparing the column densities of SiO and H$_2$. SiO abundance in gray scale with black contours
        (contour levels from 7 x 10$^{-10}$ to 47 x 10$^{-10}$ by 10 x 10$^{-10}$). The red contour shows the SiO column density at 1.5 x 10$^{13}$ cm$^{-2}$, which is equivalent to the 5$\sigma$ level and helps to estimate the uncertainty of the map. The two crosses indicate the positions of the two main continuum peaks. For more information on the abundance see Appendix \ref{A.abundance}.}         
        \label{abundance}
\end{figure}
The abundance map supports the previous findings: The observed SiO peaks are visible, and one can see that the SiO column density in the southern part is comparably low. We saw earlier that the highest SiO emission appears in the area between the two strong large scale continuum sources. This is the same for the SiO abundance, since we got the highest abundances in the middle of the region, where the H$_2$ column density is lower.\\
In different studies, it has been derived and observed that cold and quiescent gas has a SiO abundance on the order of 10 x 10$^{-12}$ \citep[e.g.,][]{Lefloch1998,RequenaTorres2007,Gerner2014}. Taking this into consideration, as well as the values in the abundance map (with a maximum of 6.8 x 10$^{-9}$, for the values see Table \ref{T.abundance}), it is clear that all the structures in Fig. \ref{abundance} show a highly increased abundance, which again implies shock activity. The derived abundance also lies in the range that is proposed by the simulations of \citet{Schilke1997} and \citet{Gusdorf2008a,Gusdorf2008b}. It is necessary to mention here that the abundance map should be viewed with some caution. To derive the H$_2$ column density, we calculated the H$^{13}$CO$^+$ column density from the data and used a N(H$^{13}$CO$^+$)/N(H$_2$) abundance, derived by \citet{Gerner2014}, in order to calculate the H$_2$ column density (for more details see Appendix \ref{A.abundance}). Therefore, there are uncertainties due to the different excitation properties of H$^{13}$CO$^+$ and SiO and due to potential differences in the H$^{13}$CO$^+$ and H$_2$ column densities. Nevertheless, the map shows where SiO is present and provides an indication of the abundance of SiO.

\section{Discussion}\label{dis}
To analyze the results of the observation, it is necessary to have a look at the physical background of SiO emission. As mentioned before, Si in quiescent gas is most common in silicates bound to dust grains. We observed a highly increased abundance of SiO, which shows that there are processes that release the Si from the dust grains into the gas phase. This is an effect of grain sputtering or vaporization in shock waves, which occur for example in protostellar outflows. Simulations concerning the emission of SiO in interstellar shocks were first conducted by \citet{Schilke1997} and extended in more recent studies by \citet{Gusdorf2008a,Gusdorf2008b}.  For the release of Si by shocks, the composition of the original dust grains plays an important role. When the Si is mostly found in the core of the grains, a shock speed higher than 25 km s$^{-1}$ is needed to release the Si. In this case, the Si is released by sputtering with heavy neutral species such as CO, and there is an additional chemical delay because of the conversion from Si to SiO \citep{Gusdorf2008a}. However, if SiO is present in the grain mantles, as discussed by \citet{Gusdorf2008b}, it can be sputtered by light and abundant species (H$_2$, He, H), and a lower shock speed is required to release the SiO from the grains. \citet{Gusdorf2008a,Gusdorf2008b} also came to the conclusion that the emission lines in the second case are broader than in the case with Si solely in the grain cores. In addition, the SiO intensity increases with SiO in the grain mantle. 
\citet{Anderl2013} deepened the studies by also considering the vaporization in grain-grain processing as a further possibility to release Si. All three studies came to the result that high SiO intensity can only be achieved by SiO in the grain mantles. Broad emission lines, as we observed them, are most likely an effect of SiO in the mantles or several unresolved shock fronts instead of a single planar shock wave. 

In addition to the processes that release SiO into the gas phase, it is also important to discuss the properties of the responsible shock. Shocks are divided into three types: jump shocks \citep[J-Type,][]{Hollenbach1979}, continuous shocks \citep[C-Type,][]{Draine1980}, and a mixture of both, which is a C-type shock with an introduced discontinuity \citep[CJ-Type,][]{Chieze1998}. The type of shock depends mainly on the presence of a magnetic field, since a magnetic field is necessary for the shock condition to become continuous, the ionization of the preshock gas, since only charged particles couple to the magnetic field, and on the velocity of the shock, since J-type shocks can have higher velocities than C-type shocks \citep{Flower2003,ISM,Anderl2013}. A main difference between the types is that J shocks lead to partial or complete dissociation of molecules, such as molecular hydrogen. Because we observed a lot of H$_2$ emission (see Fig. \ref{sio_co}), pure J shocks are fairly improbable in this case, since they would dissociate the molecular hydrogen. Studies by \citet{Gusdorf2008b} have shown that even CJ-type shocks at a velocity of 25 km s$^{-1}$ are energetic enough to dissociate the ambient hydrogen. However, in a more recent paper, \citet{Gusdorf2015} have found that CJ-type shocks reproduce H$_2$ observations even better. They found that C-type shocks are not able to reproduce their H$_2$ excitation temperatures. In conclusion, the strong observed H$_2$ emission suggests that we are not observing J-type shocks; however, we cannot differentiate further between C-type or CJ-type shocks with the present data. Another important parameter of the shocks is the velocity. However, it is very difficult to estimate the shock velocity in the current case. The SiO signal in the line wings is present until 15 km s$^{-1}$ apart from the rest velocity. This is most likely not the maximum shock velocity, as the strong SiO signal at ambient velocity indicates jets or outflows with a large velocity component in the plane of the sky.
In conclusion, this means that SiO emission in this case can be the effect of shocks with comparably high velocities (v$_{\text{shock}}>$ 25 km s$^{-1}$), because the expected C-type or CJ-type shocks can reach very high velocities, depending on the preshock density \citep{Bourlot2002}. These shocks would most likely be a result of the ongoing star formation. However, low-velocity shocks are also able to release the SiO if it is present in the core mantles.

Considering the observations now, it is quite clear that there is ongoing dust processing in the region, since the SiO abundance is highly increased. In star-forming regions, such an increased abundance is usually assumed to identify high-velocity shocks produced by protostellar jets. Additionally, we observed, in comparison to the ambient gas traced by H$^{13}$CO$^+$, broadened SiO emission lines, which are a further hint of gas movement due to jets and outflows. Especially in the northern part of the region, we found some agreement between the proposed outflows and the SiO emission at ambient velocity. This shows that the observed outflows lie in the sky plane, at least partially, which can also explain the narrow emission lines, in comparison with other high mass star-forming
regions. Especially the strong SiO peaks in the northern region seem to be closely correlated with H$_2$ peaks. This strengthens the idea of C-type shocks as the origin of the SiO emission.

However, there are also some observations that are harder to explain. The general point is that the SiO emission is very different to the high velocity CO emission. The initial idea behind the observation was to get the comparison of the high-velocity CO and the SiO data. We expected to see the the jets and outflows in both molecules; however, the SiO is less influenced by environmental gas. Therefore, it should trace gas closer to the jets and outflows and deliver more detailed information about them \citep[see, e.g.,][]{Gueth1999,Palau2006}. But the outflow features, which are clearly present in the CO data, are only represented by the SiO data in the northern part of the region. In the other parts,  the SiO emission shows a very different behavior than the CO emission. This is especially apparent in two main points.

\noindent
First of all, there is very little SiO emission in the southern area of the region. Despite strong H$_2$ emission, which shows shock activity, and several mm sources (see Fig. \ref{sio_co}), which are often identified with forming stars, the SiO abundance in this area is very low. 

\noindent
A second unexpected point of the SiO abundance is that the highest abundance occurs between the two main dust clumps, a little apart from the continuum peaks, without strong evidence of shocks or high energetic activities, such as H$_2$ emission or mm sources. Additionally, the emission has no counterpart in the high-velocity CO data, where jets and outflows should be visible. The emission line at this position (Point 4) is comparably thin and not shifted with respect to the rest velocity. These observations cannot be explained directly by the discussed theory. However, there are some possible explanations for the different observations that we want to discuss in the following. 

The first observation to discuss here is the weak SiO emission around the southern continuum peak. Earlier observations revealed a lot of mm-sources, strong H$_2$ and CO emission in this area. However, there is nearly no SiO emission around this region. This could be an effect of geometry and the different excitation properties. The observed H$_2$ and CO transitions need lower densities to be excited than the SiO (2-1) line. Moreover, the dust absorption of the 2.12 $\mu$m hydrogen emission is very high, which could mean that the H$_2$ emission emanate from the outer parts of the main core, because the emission would otherwise be absorbed by the dust. This also explains why the strong H$_2$ emission is distributed around the main continuum peak. Therefore, it could be a possibility that the outflows broke out of the cloud into less dense areas, where the density is too low for SiO to get excited. 

Another consideration is introduced by \citet{Gusdorf2008a}. When the Si is mainly found in the cores of the dust grains in form of silicates, the conversion into SiO will introduce a time delay. The magnitude of the delay depends on the shock speed and the preshock gas density (n$_H$). \citet{Gusdorf2008a} find an almost instantaneous conversion for high shock speed (v$_{\text{shock}}>$ 30 km s$^{-1}$) and high gas densities (n$_H$ = 10$^6$ cm$^{-3}$). The time delay increases with lower shock speeds and densities up to a delay of approximately 200 years for v$_{\text{shock}}=$ 25 km s$^{-1}$ and n$_H$ = 10$^6$ cm$^{-3}$. In our case, this should not be hugely important, because \citet{Beuther2002d} found a dynamical timescale for the outflows of 20000 years, and the CO data also suggest evolved outflows \citep{Beuther2003}. However, \citet{Gusdorf2008a} also find that the SiO intensity decreases with the density of the cloud. This is because in high densities, OH and O$_2$, which are necessary to convert Si into SiO, are destroyed in the shock wave. Consequently, it is possible that Si is present in the gas phase, but is not visible as SiO. This could be a possible explanation for the very dense southern regions. However, it would only be possible if the Si is solely in the grain cores and not in the mantles. The Si in the mantles is already in the form of SiO and
therefore needs no conversion. Because of the high SiO abundance in the northern part, we generally expect Si in the grain cores and mantles, which argues against this idea.

The second unexpected result was the high SiO abundance apart from the main dust emission, found between the two main gas clumps, without a counterpart in the high-velocity CO data, and accompanied by comparably thin emission lines at rest velocity. An usual explanation for comparably narrow SiO emission lines at ambient velocity is the idea that they trace postshock material, which is enriched with SiO thanks to the shock and which has been decelerated and mixed with the ambient gas \citep{Lefloch1998,Gusdorf2008a}. Depending on the dynamics of the postshock gas, the lines are broader if the gas still has a high velocity or the lines are narrower and have ambient velocity if the deceleration is almost complete. However, the observed SiO abundance for decelerated outflows is usually significantly smaller than for broad components (\citet{Codella1999,Jimenez2004} found abundances of around 10$^{-11}$-10$^{-10}$). 

This abundance value is much lower than in our case, and \citet{Gusdorf2008a} explain it as the mixing of postshock gas with ambient gas. This interpretation
also fits the non-detection in the high-velocity CO data, since the dynamics in the decelerated gas are lower. Interesting for the strength of the deceleration is the age of the outflow and the expected timescale. \citet{Beuther2002d} found dynamical timescales of around 20000 years. This is quite large on the timescale of outflows, which means that the outflows are already evolved, and the idea of decelerated postshock gas is probable. However, the dynamical timescale is only a rough estimate, since \citet{Beuther2002d} calculated it based on the assumption of two large outflows. 

Our main candidate for emission due to decelerated postshock gas is the emission peak around Point 4. This is one of the narrowest emission lines (FWHM = 6.9 km s$^{-1}$), and it has no increased signal in the line wings, and it peaks at rest velocity. However, the line is still clearly broadened in comparison to the corresponding H$^{13}$CO$^+$ line (FWHM = 1.8 km s$^{-1}$, see Point 4 in Table \ref{table_spek}). Also, at this point, the SiO abundance is similar to the other points with broader lines and not two or three magnitudes smaller, as expected from \citet{Gusdorf2008a} for gas that has been decelerated in ambient gas. One explanation could be that the deceleration of the gas is caused by colliding outflows that emanate from the northern and southern continuum peaks. This would explain the emission line at ambient velocity and also the location of the emission between the two main clumps. It also explains the high SiO abundance, since the two colliding flows both contain SiO-enriched postshock gas. Therefore, two or more colliding outflows can be an explanation for the emission feature around Point 4.

Another possibility for this emission was discussed by \citet{Jimenez2010}. They observed a similar feature with a strong emission line offset from the main continuum signal and explained it by outflows from a nearby evolving protostar. This would be an explanation on a smaller scale than the colliding outflows emanating from the main regions. There are two NIR observations of the region by \citet{Beuther2002b} and by \citet{Martin-Hernandez2008}. However, the more recent observation has a different field of view, which is why we show the data from \citet{Beuther2002b} in Fig. \ref{sio_co}. There are two faint sources a little north of Point 4, which have to be taken into consideration in this case. However, both sources are very weak, and \citet{Martin-Hernandez2008} only observed them in the K Band and in the shorter wavelength of the Spitzer data (at 3.6 and 4.5 $\mu$m). The weakness of the sources and the non-detection in longer Spitzer data, where one would expect a strong signal for young protostars, makes it rather unlikely that these sources are evolving protostars and that they cause the highly increased SiO abundance. Therefore, the theory of colliding outflows is more plausible in this case.

So far, we have only considered high-velocity shocks due to protostellar jets. However, there are some studies that also considered the possibility of SiO emission due to low-velocity shocks ($\leq$ 10 km s$^{-1}$). This is a possibility that could explain why the SiO emission lines are rather narrow and only slightly shifted with respect to the velocity of rest. \citet{NguyenLuong2013} and \citet{Jimenez2010} have introduced the possibility of low-velocity shocks, which are most likely an effect of large converging flows, such as cloud collision or interaction between two filaments. \citet{Jimenez2010} consider this possibility for emissions apart from the high-density structures. This could also be a valid possibility for our region, since the spectra from W43 \citep{NguyenLuong2013} are, in view of line width and the velocity shift of the SiO line, comparable to those from IRAS 19410 \& 2336. Additionally, the shift in the line peak in the present data lies with a maximum of 6.1 km s$^{-1}$ (see Table \ref{table_spek}) far below the predicted shock velocity for high-velocity shocks (v$_{\text{shock}}>$ 25 km s$^{-1}$, \citet{Gusdorf2008a,Gusdorf2008b}). This could be a hint of low-velocity shocks, but it could, as mentioned, also be an effect of the projection in the observed plane and the acceleration of environmental gas.

\noindent The main reason for \citet{NguyenLuong2013} considering low-velocity shocks due to large colliding flows was the fact that they observed large SiO structures with relatively uniform emission  over the whole area. Again, this is in contrast to our observation that the whole southern part of IRAS 19410 \& 2336 lacks SiO emission.  Therefore, a collision of flows could again explain the SiO emission in the middle of the region. Still, it cannot explain the only partially present SiO emission in the south or the differences between SiO and CO data. Furthermore, the detection of H$_2$ emission as a jet shock indicator in the south without any SiO counterpart could not be explained by this.

\section{Conclusion}
After combining the different results, we found that the SiO emission is most likely caused by high-velocity C-type or CJ-type shocks probably from protostellar jets. This explains the emission in the northern part, as well as the observed line widths. The emission in the middle of the region could be a result of decelerated postshock gas due to two or more colliding outflows. 
The only part of the observation that does not fit  this explanation is the missing SiO emission in the southern region. The observed H$_2$ emission clearly shows shock activity, and the CO data exhibit several bipolar structures. The most valid explanation for us is that there is ongoing jet and outflow activity in the southern part, which is not visible in the SiO data. The missing SiO emission could be a result of the different excitation properties of the molecules and the shock geometry. When the shocks break out of the densest area, as the H$_2$ data suggest, the density could be too low for SiO to get excited. However, the opposite explanation that SiO is also not visible owing to densities that are too high in the core has to be considered here. 

Our final conclusion from this observation is that SiO emission has to be used with some caution, because a lot of effects can have an influence on the emission. This is nicely shown by the observed region, since it can be divided into three different parts, which all show a completely different SiO morphology. Especially the southern area of IRAS 19410+2336 leads to the conclusion that SiO should only be used as a shock tracer in the sense that if one observes SiO emission, it is most likely an effect of shocks. However, the reverse argument that a non-detection of SiO also means no existing shocks is certainly not valid because several different effects can influence the SiO emission.

\bibliography{mylib2}  

\begin{thebibliography}{53}
\expandafter\ifx\csname natexlab\endcsname\relax\def\natexlab#1{#1}\fi

\bibitem[{{Anderl} {et~al.}(2013){Anderl}, {Guillet}, {Pineau des For{\^e}ts},
  \& {Flower}}]{Anderl2013}
{Anderl}, S., {Guillet}, V., {Pineau des For{\^e}ts}, G., \& {Flower}, D.~R.
  2013, \aap, 556, A69

\bibitem[{{Arce} {et~al.}(2007){Arce}, {Shepherd}, {Gueth}, {Lee}, {Bachiller},
  {Rosen}, \& {Beuther}}]{Arce2007}
{Arce}, H.~G., {Shepherd}, D., {Gueth}, F., {et~al.} 2007, Protostars and
  Planets V, 245

\bibitem[{{Beuther} {et~al.}(2002{\natexlab{a}}){Beuther}, {Kerp}, {Preibisch},
  {Stanke}, \& {Schilke}}]{Beuther2002c}
{Beuther}, H., {Kerp}, J., {Preibisch}, T., {Stanke}, T., \& {Schilke}, P.
  2002{\natexlab{a}}, \aap, 395, 169

\bibitem[{{Beuther} \& {Schilke}(2004)}]{Beuther2004}
{Beuther}, H. \& {Schilke}, P. 2004, Science, 303, 1167

\bibitem[{{Beuther} {et~al.}(2002{\natexlab{b}}){Beuther}, {Schilke}, {Gueth},
  {McCaughrean}, {Andersen}, {Sridharan}, \& {Menten}}]{Beuther2002b}
{Beuther}, H., {Schilke}, P., {Gueth}, F., {et~al.} 2002{\natexlab{b}}, \aap,
  387, 931

\bibitem[{{Beuther} {et~al.}(2002{\natexlab{c}}){Beuther}, {Schilke}, {Menten},
  {Motte}, {Sridharan}, \& {Wyrowski}}]{Beuther2002}
{Beuther}, H., {Schilke}, P., {Menten}, K.~M., {et~al.} 2002{\natexlab{c}},
  \apj, 566, 945

\bibitem[{{Beuther} {et~al.}(2005){Beuther}, {Schilke}, {Menten}, {Motte},
  {Sridharan}, \& {Wyrowski}}]{Beuther2005}
{Beuther}, H., {Schilke}, P., {Menten}, K.~M., {et~al.} 2005, \apj, 633, 535

\bibitem[{{Beuther} {et~al.}(2002{\natexlab{d}}){Beuther}, {Schilke},
  {Sridharan}, {Menten}, {Walmsley}, \& {Wyrowski}}]{Beuther2002d}
{Beuther}, H., {Schilke}, P., {Sridharan}, T.~K., {et~al.} 2002{\natexlab{d}},
  \aap, 383, 892

\bibitem[{{Beuther} {et~al.}(2003){Beuther}, {Schilke}, \&
  {Stanke}}]{Beuther2003}
{Beuther}, H., {Schilke}, P., \& {Stanke}, T. 2003, \aap, 408, 601

\bibitem[{{Beuther} \& {Sridharan}(2007)}]{Beuther2007a}
{Beuther}, H. \& {Sridharan}, T.~K. 2007, \apj, 668, 348

\bibitem[{{Bonnell}(2008)}]{Bonnell2008}
{Bonnell}, I.~A. 2008, in Astronomical Society of the Pacific Conference
  Series, Vol. 390, Pathways Through an Eclectic Universe, ed. J.~H. {Knapen},
  T.~J. {Mahoney}, \& A.~{Vazdekis}, 26

\bibitem[{{Bonnell} {et~al.}(2007){Bonnell}, {Larson}, \&
  {Zinnecker}}]{Bonnell2007}
{Bonnell}, I.~A., {Larson}, R.~B., \& {Zinnecker}, H. 2007, Protostars and
  Planets V, 149

\bibitem[{{Cesaroni} {et~al.}(1999){Cesaroni}, {Felli}, {Jenness}, {Neri},
  {Olmi}, {Robberto}, {Testi}, \& {Walmsley}}]{Cesaroni1999}
{Cesaroni}, R., {Felli}, M., {Jenness}, T., {et~al.} 1999, \aap, 345, 949

\bibitem[{{Chieze} {et~al.}(1998){Chieze}, {Pineau des Forets}, \&
  {Flower}}]{Chieze1998}
{Chieze}, J.-P., {Pineau des Forets}, G., \& {Flower}, D.~R. 1998, \mnras, 295,
  672

\bibitem[{{Codella} {et~al.}(1999){Codella}, {Bachiller}, \&
  {Reipurth}}]{Codella1999}
{Codella}, C., {Bachiller}, R., \& {Reipurth}, B. 1999, \aap, 343, 585

\bibitem[{{Codella} {et~al.}(2013){Codella}, {Beltr{\'a}n}, {Cesaroni},
  {Moscadelli}, {Neri}, {Vasta}, \& {Zhang}}]{Codella2013}
{Codella}, C., {Beltr{\'a}n}, M.~T., {Cesaroni}, R., {et~al.} 2013, \aap, 550,
  A81

\bibitem[{{Codella} {et~al.}(2007){Codella}, {Cabrit}, {Gueth}, {Cesaroni},
  {Bacciotti}, {Lefloch}, \& {McCaughrean}}]{Codella2007}
{Codella}, C., {Cabrit}, S., {Gueth}, F., {et~al.} 2007, \aap, 462, L53

\bibitem[{{Draine}(1980)}]{Draine1980}
{Draine}, B.~T. 1980, \apj, 241, 1021

\bibitem[{{Flower} \& {Pineau des For{\^e}ts}(2003)}]{Flower2003}
{Flower}, D.~R. \& {Pineau des For{\^e}ts}, G. 2003, \mnras, 341, 1272

\bibitem[{{Gerner} {et~al.}(2014){Gerner}, {Beuther}, {Semenov}, {Linz},
  {Vasyunina}, {Bihr}, {Shirley}, \& {Henning}}]{Gerner2014}
{Gerner}, T., {Beuther}, H., {Semenov}, D., {et~al.} 2014, \aap, 563, A97

\bibitem[{{Gueth} \& {Guilloteau}(1999)}]{Gueth1999}
{Gueth}, F. \& {Guilloteau}, S. 1999, \aap, 343, 571

\bibitem[{{Guilloteau} {et~al.}(1992){Guilloteau}, {Delannoy}, {Downes},
  {Greve}, {Guelin}, {Lucas}, {Morris}, {Radford}, {Wink}, {Cernicharo},
  {Forveille}, {Garcia-Burillo}, {Neri}, {Blondel}, {Perrigourad}, {Plathner},
  \& {Torres}}]{Guilloteau1992}
{Guilloteau}, S., {Delannoy}, J., {Downes}, D., {et~al.} 1992, \aap, 262, 624

\bibitem[{{Gusdorf} {et~al.}(2008{\natexlab{a}}){Gusdorf}, {Cabrit}, {Flower},
  \& {Pineau Des For{\^e}ts}}]{Gusdorf2008a}
{Gusdorf}, A., {Cabrit}, S., {Flower}, D.~R., \& {Pineau Des For{\^e}ts}, G.
  2008{\natexlab{a}}, \aap, 482, 809

\bibitem[{{Gusdorf} {et~al.}(2008{\natexlab{b}}){Gusdorf}, {Pineau Des
  For{\^e}ts}, {Cabrit}, \& {Flower}}]{Gusdorf2008b}
{Gusdorf}, A., {Pineau Des For{\^e}ts}, G., {Cabrit}, S., \& {Flower}, D.~R.
  2008{\natexlab{b}}, \aap, 490, 695

\bibitem[{{Gusdorf} {et~al.}(2015){Gusdorf}, {Riquelme}, {Anderl},
  {Eisl{\"o}ffel}, {Codella}, {G{\'o}mez-Ruiz}, {Graf}, {Kristensen},
  {Leurini}, {Parise}, {Requena-Torres}, {Ricken}, \&
  {G{\"u}sten}}]{Gusdorf2015}
{Gusdorf}, A., {Riquelme}, D., {Anderl}, S., {et~al.} 2015, \aap, 575, A98

\bibitem[{{Hezareh} {et~al.}(2008){Hezareh}, {Houde}, {McCoey}, {Vastel}, \&
  {Peng}}]{Hezareh2008}
{Hezareh}, T., {Houde}, M., {McCoey}, C., {Vastel}, C., \& {Peng}, R. 2008,
  \apj, 684, 1221

\bibitem[{{Hollenbach} \& {McKee}(1979)}]{Hollenbach1979}
{Hollenbach}, D. \& {McKee}, C.~F. 1979, \apjs, 41, 555

\bibitem[{{Jim{\'e}nez-Serra} {et~al.}(2010){Jim{\'e}nez-Serra}, {Caselli},
  {Tan}, {Hernandez}, {Fontani}, {Butler}, \& {van Loo}}]{Jimenez2010}
{Jim{\'e}nez-Serra}, I., {Caselli}, P., {Tan}, J.~C., {et~al.} 2010, \mnras,
  406, 187

\bibitem[{{Jim{\'e}nez-Serra} {et~al.}(2009){Jim{\'e}nez-Serra},
  {Mart{\'{\i}}n-Pintado}, {Caselli}, {Viti}, \&
  {Rodr{\'{\i}}guez-Franco}}]{Jimenez2009}
{Jim{\'e}nez-Serra}, I., {Mart{\'{\i}}n-Pintado}, J., {Caselli}, P., {Viti},
  S., \& {Rodr{\'{\i}}guez-Franco}, A. 2009, \apj, 695, 149

\bibitem[{{Jim{\'e}nez-Serra} {et~al.}(2004){Jim{\'e}nez-Serra},
  {Mart{\'{\i}}n-Pintado}, {Rodr{\'{\i}}guez-Franco}, \&
  {Marcelino}}]{Jimenez2004}
{Jim{\'e}nez-Serra}, I., {Mart{\'{\i}}n-Pintado}, J.,
  {Rodr{\'{\i}}guez-Franco}, A., \& {Marcelino}, N. 2004, \apjl, 603, L49

\bibitem[{{Jim{\'e}nez-Serra} {et~al.}(2011){Jim{\'e}nez-Serra},
  {Mart{\'{\i}}n-Pintado}, {Winters}, {Rodr{\'{\i}}guez-Franco}, \&
  {Caselli}}]{Jimenez2011}
{Jim{\'e}nez-Serra}, I., {Mart{\'{\i}}n-Pintado}, J., {Winters}, J.-M.,
  {Rodr{\'{\i}}guez-Franco}, A., \& {Caselli}, P. 2011, \apj, 739, 80

\bibitem[{{Krumholz}(2006)}]{Krumholz2006}
{Krumholz}, M.~R. 2006, in Astronomical Society of the Pacific Conference
  Series, Vol. 352, New Horizons in Astronomy: Frank N. Bash Symposium, ed.
  S.~J. {Kannappan}, S.~{Redfield}, J.~E. {Kessler-Silacci}, M.~{Landriau}, \&
  N.~{Drory}, 31

\bibitem[{{Le Bourlot} {et~al.}(2002){Le Bourlot}, {Pineau des For{\^e}ts},
  {Flower}, \& {Cabrit}}]{Bourlot2002}
{Le Bourlot}, J., {Pineau des For{\^e}ts}, G., {Flower}, D.~R., \& {Cabrit}, S.
  2002, \mnras, 332, 985

\bibitem[{{Lee} {et~al.}(2007){Lee}, {Ho}, {Hirano}, {Beuther}, {Bourke},
  {Shang}, \& {Zhang}}]{Lee2007}
{Lee}, C.-F., {Ho}, P.~T.~P., {Hirano}, N., {et~al.} 2007, \apj, 659, 499

\bibitem[{{Lefloch} {et~al.}(1998){Lefloch}, {Castets}, {Cernicharo}, \&
  {Loinard}}]{Lefloch1998}
{Lefloch}, B., {Castets}, A., {Cernicharo}, J., \& {Loinard}, L. 1998, \apjl,
  504, L109

\bibitem[{{Leurini} {et~al.}(2013){Leurini}, {Codella}, {Gusdorf}, {Zapata},
  {G{\'o}mez-Ruiz}, {Testi}, \& {Pillai}}]{Leurini2013}
{Leurini}, S., {Codella}, C., {Gusdorf}, A., {et~al.} 2013, \aap, 554, A35

\bibitem[{{Leurini} {et~al.}(2014){Leurini}, {Codella}, {L{\'o}pez-Sepulcre},
  {Gusdorf}, {Csengeri}, \& {Anderl}}]{Leurini2014}
{Leurini}, S., {Codella}, C., {L{\'o}pez-Sepulcre}, A., {et~al.} 2014, \aap,
  570, A49

\bibitem[{{L{\'o}pez-Sepulcre} {et~al.}(2011){L{\'o}pez-Sepulcre}, {Walmsley},
  {Cesaroni}, {Codella}, {Schuller}, {Bronfman}, {Carey}, {Menten}, {Molinari},
  \& {Noriega-Crespo}}]{LopezSepulcre2011}
{L{\'o}pez-Sepulcre}, A., {Walmsley}, C.~M., {Cesaroni}, R., {et~al.} 2011,
  \aap, 526, L2

\bibitem[{{Mart{\'{\i}}n-Hern{\'a}ndez}
  {et~al.}(2008{\natexlab{a}}){Mart{\'{\i}}n-Hern{\'a}ndez}, {Bik}, {Puga},
  {N{\"u}rnberger}, \& {Bronfman}}]{Martin2008}
{Mart{\'{\i}}n-Hern{\'a}ndez}, N.~L., {Bik}, A., {Puga}, E., {N{\"u}rnberger},
  D.~E.~A., \& {Bronfman}, L. 2008{\natexlab{a}}, \aap, 489, 229

\bibitem[{{Mart{\'{\i}}n-Hern{\'a}ndez}
  {et~al.}(2008{\natexlab{b}}){Mart{\'{\i}}n-Hern{\'a}ndez}, {Bik}, {Puga},
  {N{\"u}rnberger}, \& {Bronfman}}]{Martin-Hernandez2008}
{Mart{\'{\i}}n-Hern{\'a}ndez}, N.~L., {Bik}, A., {Puga}, E., {N{\"u}rnberger},
  D.~E.~A., \& {Bronfman}, L. 2008{\natexlab{b}}, \aap, 489, 229

\bibitem[{{McKee} \& {Tan}(2003)}]{Mckee2003}
{McKee}, C.~F. \& {Tan}, J.~C. 2003, \apj, 585, 850

\bibitem[{{Motte} {et~al.}(2007){Motte}, {Bontemps}, {Schilke}, {Schneider},
  {Menten}, \& {Brogui{\`e}re}}]{Motte2007}
{Motte}, F., {Bontemps}, S., {Schilke}, P., {et~al.} 2007, \aap, 476, 1243

\bibitem[{{Nguyen-Lu'o'ng} {et~al.}(2013){Nguyen-Lu'o'ng}, {Motte}, {Carlhoff},
  {Louvet}, {Lesaffre}, {Schilke}, {Hill}, {Hennemann}, {Gusdorf}, {Didelon},
  {Schneider}, {Bontemps}, {Duarte-Cabral}, {Menten}, {Martin}, {Wyrowski},
  {Bendo}, {Roussel}, {Bernard}, {Bronfman}, {Henning}, {Kramer}, \&
  {Heitsch}}]{NguyenLuong2013}
{Nguyen-Lu'o'ng}, Q., {Motte}, F., {Carlhoff}, P., {et~al.} 2013, \apj, 775, 88

\bibitem[{{Palau} {et~al.}(2006){Palau}, {Ho}, {Zhang}, {Estalella}, {Hirano},
  {Shang}, {Lee}, {Bourke}, {Beuther}, \& {Kuan}}]{Palau2006}
{Palau}, A., {Ho}, P.~T.~P., {Zhang}, Q., {et~al.} 2006, \apjl, 636, L137

\bibitem[{{Requena-Torres} {et~al.}(2007){Requena-Torres}, {Marcelino},
  {Jim{\'e}nez-Serra}, {Mart{\'{\i}}n-Pintado}, {Mart{\'{\i}}n}, \&
  {Mauersberger}}]{RequenaTorres2007}
{Requena-Torres}, M.~A., {Marcelino}, N., {Jim{\'e}nez-Serra}, I., {et~al.}
  2007, \apjl, 655, L37

\bibitem[{{Rod{\'o}n} {et~al.}(2012){Rod{\'o}n}, {Beuther}, \&
  {Schilke}}]{Rodon2012}
{Rod{\'o}n}, J.~A., {Beuther}, H., \& {Schilke}, P. 2012, \aap, 545, A51

\bibitem[{{Schilke} {et~al.}(1997){Schilke}, {Walmsley}, {Pineau des Forets},
  \& {Flower}}]{Schilke1997}
{Schilke}, P., {Walmsley}, C.~M., {Pineau des Forets}, G., \& {Flower}, D.~R.
  1997, \aap, 321, 293

\bibitem[{{Sridharan} {et~al.}(2002){Sridharan}, {Beuther}, {Schilke},
  {Menten}, \& {Wyrowski}}]{Sridharan2002}
{Sridharan}, T.~K., {Beuther}, H., {Schilke}, P., {Menten}, K.~M., \&
  {Wyrowski}, F. 2002, \apj, 566, 931

\bibitem[{{Tan} {et~al.}(2014){Tan}, {Beltran}, {Caselli}, {Fontani}, {Fuente},
  {Krumholz}, {McKee}, \& {Stolte}}]{Tan2014}
{Tan}, J.~C., {Beltran}, M.~T., {Caselli}, P., {et~al.} 2014, Protostars and
  Planets VI, 149

\bibitem[{Tielens(2005)}]{ISM}
Tielens, A. G. G.~M. 2005, The Physics and Chemistry of the Interstellar Medium
  (Cambridge University Press)

\bibitem[{{Vasyunina} {et~al.}(2011){Vasyunina}, {Linz}, {Henning},
  {Zinchenko}, {Beuther}, \& {Voronkov}}]{Vasyunina2011}
{Vasyunina}, T., {Linz}, H., {Henning}, T., {et~al.} 2011, \aap, 527, A88

\bibitem[{{Xu} {et~al.}(2009){Xu}, {Reid}, {Menten}, {Brunthaler}, {Zheng}, \&
  {Moscadelli}}]{Xu2009}
{Xu}, Y., {Reid}, M.~J., {Menten}, K.~M., {et~al.} 2009, \apj, 693, 413

\bibitem[{{Zinnecker} \& {Yorke}(2007)}]{Zinnecker2007}
{Zinnecker}, H. \& {Yorke}, H.~W. 2007, \araa, 45, 481

\end{thebibliography}
\bibliographystyle{aa}

\appendix

\section{SiO Abundance}\label{A.abundance}
\begin{figure}[htb]
        \centering
        \includegraphics[width=0.45\textwidth]{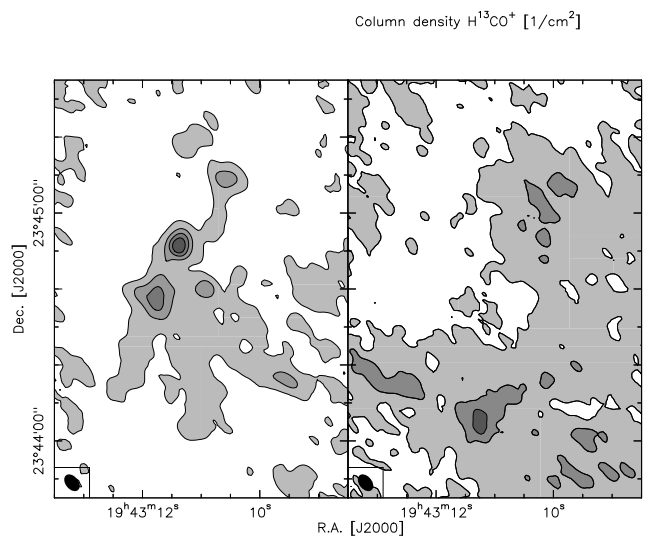}
         \caption{Column density of SiO (left) and H$^{13}$CO$^+$ (right) calculated at T=40 K. The contour levels of the SiO image are 15(30)60 x 10$^{12}$ cm$^{-2}$ and for the H$^{13}$CO$^+$ image 13(13)40 x 10$^{12}$ cm$^{-2}$. The lowest contour in both cases shows the 5$\sigma$ level.}
                \label{column}
\end{figure}

The abundance map of SiO is calculated from the column density maps of H$^{13}$CO$^+$ and SiO. To calculate the column densities from the emission lines, we used the calculations from \citet{Hezareh2008}:
\begin{equation}
N_u=\frac{8\pi k \nu^2}{hc^3 A_{ul}}\frac{\tau}{(1-e^{-\tau})} \int T_b d\nu
\label{column-dens1}
.\end{equation}
The total column density can then be calculated with
\begin{equation}
N_{tot}=N_u\frac{U(T_{ex})}{g_u}exp\left(\frac{E_u}{kT_{ex}}\right)
\label{column-dens2}
.\end{equation}
In the equation, A$_{ul}$ is the Einstein coefficient for a given transition with g$_u$ and E$_u$/k, the upper state degeneracy and temperature with the Boltzmann constant k. Here, T$_{ex}$ is the excitation temperature and U(T$_{ex}$) is the partition function at this temperature, and T$_b$ is the brightness temperature, which we derive from the data. In the calculations, we assume that the excitation temperature is equal to the kinetic temperature, which should be approximately correct at the given high densities. As temperature, 40 K is used \citep{Rodon2012} and $\tau$ is the optical depth. For the assumption of optically thin lines holds $\tau \ll 1,$ and the factor $\tau/(1-e^{-\tau})$ approaches unity. This assumption is probably not true for the denser part of the SiO map. This means that the map only gives a lower limit for the column density, especially at the densest points. For more information on the column density and a derivation of the used formulas, see \citet{Hezareh2008} and \citet{ISM}. \\
The column density maps calculated from the H$^{13}$CO$^+$ and SiO data can be seen in Fig. \ref{column}

\begin{figure}[h]
\centering
\includegraphics[width=0.4\textwidth]{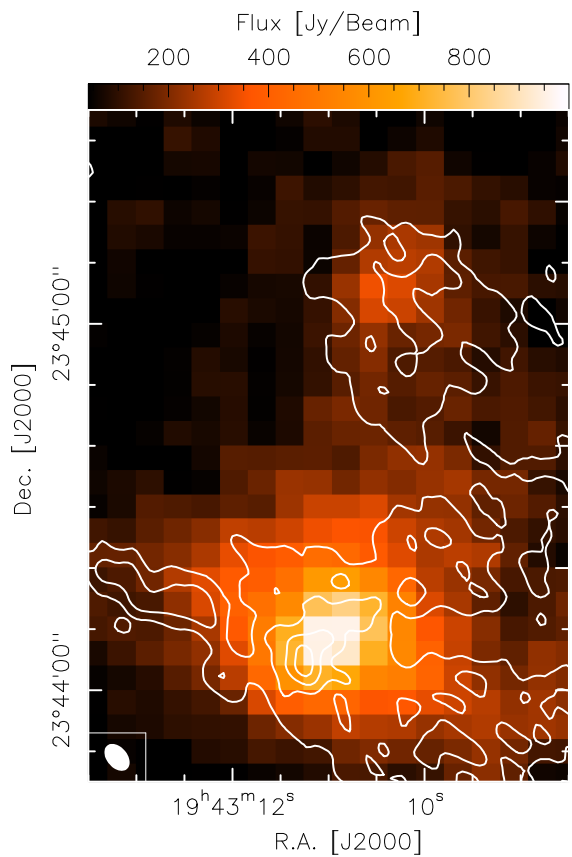}
\caption{Line emission from H$^{13}$CO$^+$ shown as white lines (levels are 0.4(0.2)2 Jy beam$^{-1}$) with the 1.2 mm continuum data from \citet{Beuther2002} in the background.}
\label{dense-gas}
\end{figure}

With the help of Fig. \ref{column}, it is possible to calculate the abundance of SiO with respect to H$_2$. Therefore, it is necessary to know the column density of H$_2$. At this point we assume that the distribution of H$_2$ is roughly the same as the one from H$^{13}$CO$^+$. This assumption is reasonable, as it can be seen in Fig. \ref{dense-gas}. The figure shows that the H$^{13}$CO$^+$ distribution describes the dust continuum emission well, which is directly related to the H$_2$ distribution. To get the column density of H$_2$, we only need the H$^{13}$CO$^+$ to H$_2$ abundance. This information can be obtained from  \citet{Gerner2014}. They investigated several young massive star formation regions, including IRAS 19410+2336, and calculated several abundances. For high-mass protostellar objects (HMPO), which is the evolutionary state of IRAS 19410+2336, they came to the following result:
\begin{equation}
\frac{N(H^{13}CO^+)}{N(H_2)}=4.5\cdot 10^{-10}
\label{equ_abudnance}
.\end{equation}
The final abundance map is the quotient of the column density maps from SiO and H$^{13}$CO$^+$ multiplied with the aforementioned abundance factor. This leads to the N(SiO)/N(H$_2$) abundance map, which can be seen in Fig. \ref{abundance}.
\section{Tables}
\newcolumntype{Y}{>{\centering\arraybackslash}X}                        
\begin{table*}
\caption{Properties of the spectral lines in Fig. \ref{spektren1} with the errors in brackets }
\label{table_spek}
\vspace{3pt}
\centering
\begin{threeparttable}
\begin{tabularx}{0.95\textwidth}{c YY YY YY}
\hline\hline 
\multirow{2}{*}{Point} & \multicolumn{2}{c}{Position (J2000.0)} &   \multicolumn{2}{c}{\rule{0pt}{0.5cm} Peak velocity [km s$^{-1}$]}& \multicolumn{2}{c}{Line width [km s$^{-1}$]} \\ \cmidrule(l{10pt}r{10pt}){2-3} \cmidrule(l{15pt}r{15pt}){4-5} \cmidrule(l{15pt}r{15pt}){6-7}
& R.A  & Dec. & SiO \rule{0pt}{0.5cm} &H$^{13}$CO$^+$ & SiO &H$^{13}$CO$^+$ \\\hline
 \rule{0pt}{0.5cm}  1   & 19$^h$43$^m$10.6$^s$ & 23$^\circ$45'08.6"             & 16.1(0.4)       & 20.9(0.1) & 8.5(0.9)          & 2.2(0.1) \\
2       & 19$^h$43$^m$11.4$^s$ & 23$^\circ$44'51.9"             & 20.0(0.3)         & 21.7(0.2) & 10.2(0.6)         & 3.5(0.4) \\
3       & 19$^h$43$^m$10.9$^s$ & 23$^\circ$44'40.3"             & 24.3(0.4)         & 21.9(0.1) & 10.3(0.9)         & 2.0(0.2) \\
4       & 19$^h$43$^m$11.7$^s$ & 23$^\circ$44'34.3"             & 22.5(0.1)         & 22.4(0.1) & 6.9(0.3)          & 1.8(0.1) \\
5  & 19$^h$43$^m$10.7$^s$ & 23$^\circ$44'58.0"          & 18.7(0.7)     & 21.2(0.1) & 9.7(1.4)            & 2.6(0.1) \\
6       & 19$^h$43$^m$11.2$^s$ & 23$^\circ$44'02.9"     & 22.0(0.3)             & 22.3(0.1)       & 4.5(0.8)              & 2.7(0.1) 
\\\hline 
\end{tabularx}
\end{threeparttable} 
\end{table*}

\begin{table*}
\caption{Column densities and abundances at different points of the region. Points from Fig. \ref{spektren1}, column density, and abundance maps in Fig. \ref{column} \& \ref{abundance}. }
\label{T.abundance}
\vspace{3pt}
\centering
\begin{tabularx}{0.95\textwidth}{cYYYYc}
\hline\hline
\multirow{2}{*}{Point} & \multicolumn{2}{c}{Position (J2000.0)} & \multicolumn{2}{c}{\rule{0pt}{0.5cm} Column density [10$^{13}$cm$^{-1}$]}  & Abundance [10$^{-9}$] \\ \cmidrule(l{10pt}r{10pt}){2-3} \cmidrule(l{10pt}r{10pt}){4-5} \cmidrule(l{10pt}r{10pt}){6-6} 
&R.A & Dec.  \rule{0pt}{0.5cm}  &  SiO &H$^{13}$CO$^+$ &[SiO]/[H$_2$] \\\hline
1 \rule{0pt}{0.5cm}     & 19$^h$43$^m$10.6$^s$ & 23$^\circ$45'08.6"& 6.77         & 2.43  &       1.25 \\
2       & 19$^h$43$^m$11.4$^s$ & 23$^\circ$44'51.9"                     &  13.6   & 1.01  &       6.75 \\
3       & 19$^h$43$^m$10.9$^s$ & 23$^\circ$44'40.3"                             &  7.06   &       0.830   &       3.82 \\
4       & 19$^h$43$^m$11.7$^s$ & 23$^\circ$44'34.3"                     &  7.44   &       1.06            &       3.29 \\
5  & 19$^h$43$^m$10.7$^s$ & 23$^\circ$44'58.0"                  &  1.89 &       2.19            &       0.423 \\
6       & 19$^h$43$^m$11.2$^s$ & 23$^\circ$44'02.9"                     &  3.38   &       4.14            &       0.363  \\\hline 
\end{tabularx}
\end{table*}

\end{document}